\documentclass{elsart}

\usepackage{epsfig}
\usepackage{subfigure}
\usepackage{graphics}
\usepackage{amsmath}
\usepackage{adjustbox}

\begin{document}
\begin{frontmatter}
\title{Transverse momentum spectra of hadrons in $p+p$ collisions
at CERN SPS energies from the UrQMD transport model}

\author{V.~Ozvenchuk}
\ead{Vitalii.Ozvenchuk@ifj.edu.pl}
%\email{Vitalii.Ozvenchuk@ifj.edu.pl}
\and
\author{A.~Rybicki}

\address{H.~Niewodnicza\'nski Institute of Nuclear Physics, Polish Academy of Sciences, %
  Radzikowskiego 152, 31-342 Krak\'ow, %
  Poland %
}

%\affiliation{%
%  H.~Niewodnicza\'nski Institute of Nuclear Physics, Polish Academy of Sciences, %
%  Radzikowskiego 152, 31-342 Krak\'ow, %
%  Poland %
%}

\begin{abstract}
The UrQMD transport model, version 3.4, is used to study the new
experimental data on transverse momentum spectra of $\pi^{\pm}$,
$K^{\pm}$, $p$ and $\bar p$ produced in inelastic $p+p$ interactions
at SPS energies, recently published by the NA61/SHINE Collaboration.
The comparison of model predictions to these new measurements is
presented as a function of collision energy for central and forward
particle rapidity intervals. In addition, the inverse slope
parameters characterizing the transverse momentum distributions are
extracted from the predicted spectra and compared to the
corresponding values obtained from NA61/SHINE distributions, as a
function of particle rapidity and collision energy. A complex
pattern of deviations between the experimental data and the UrQMD
model emerges. For charged pions, the fair agreement visible at top
SPS energies deteriorates with the decreasing energy. For charged
$K$ mesons, UrQMD significantly underpredicts positive kaon
production at lower beam momenta. It also underpredicts the central
rapidity proton yield at top collision energy and overpredicts
antiproton production at all considered energies. We conclude that
the new experimental data analyzed in this paper still constitute a
challenge for the present version of the model.
\end{abstract}


\end{frontmatter}
%\maketitle

\section{Introduction}
\label{Introduction} Recently the NA61/SHINE Collaboration published
new, detailed experimental results~\cite{NA61_pp} on inclusive
spectra and mean multiplicities of $\pi^{\pm}$, $K^{\pm}$, $p$ and
$\bar p$ produced in inelastic $p+p$ interactions at 20, 31, 40, 80
and 158 GeV/$c$ at the CERN Super Proton Synchrotron (SPS). These
measurements were meant as a baseline in the study of the properties
of the onset of deconfinement and the possibility to observe the
critical point of strongly-interacting matter in nucleus-nucleus
collisions. As such they were motivated by the observation of the
onset of
deconfinement~\cite{NA49_deconfinement_1,NA49_deconfinement_2} in
central $Pb+Pb$ reactions at about 30 GeV/$c$ by the NA49
Collaboration at the CERN SPS (for comparison, see also the results
obtained by the RHIC beam energy scan
programme~\cite{RHIC_deconfinement} and by the
LHC~\cite{LHC_deconfinement}).

This new baseline significantly extends the present experimental
knowledge on the energy dependence of particle production in $p+p$
reactions at the SPS, where high quality data of comparable
phase-space coverage existed only from the NA49 detector at top SPS
energy~\cite{na49-pp-pion,na49-pp-proton,na49-pp-kaon}. The
importance of these new results for the understanding on nuclear
collisions has been many times advertised by the NA61/SHINE
Collaboration (see e.g. results on kaon over pion multiplicity
ratios presented in~\cite{Aduszkiewicz-qm2017}). On the other hand,
they also refine the basis of our knowledge on inclusive soft
particle production in general, a subject of renewed interest in
view of the recent findings made e.g. in $p+p$ collisions at the
LHC~\cite{alice-nature2017}. For both above reasons, it seems that
an overall assessment on how these new data are being described by
the presently available theoretical models of soft hadronic
collisions has its own importance for the whole field of hadronic
and high energy nuclear physics.

In the present paper, we perform the comparison of the recently
obtained NA61/SHINE experimental data to the UrQMD transport model.
We analyze the transverse momentum spectra of $\pi^{\pm}$,
$K^{\pm}$, $p$ and $\bar p$ produced in inelastic $p+p$ interactions
at 20, 31, 40, 80 and 158 GeV/$c$ and extract the inverse slope
parameters as a function of the particle rapidity and collision
energy. The inverse slopes are obtained from fits to the transverse
momentum spectra of the different particles.

We note that earlier analyses have been performed in
Refs.~\cite{Vovchenko} and~\cite{Uzh}, where the authors have used
the UrQMD transport model to study the precedent NA61/SHINE dataset
including only spectra of negatively charged pions produced in
inelastic $p+p$ collisions at 20, 31, 40, 80 and 158
GeV/$c$~\cite{NA61_pp_previous}, as well as NA49 data on
$\pi^{\pm}$, $K^{\pm}$, $p$ and $\bar p$ spectra in $p+p$
interactions only at top SPS
energy~\cite{na49-pp-pion,na49-pp-proton,na49-pp-kaon}. In addition,
in Ref.~\cite{Uzh} possible improvements of the UrQMD model have
been studied and we will discuss them in more detail in
Sec.~\ref{Summary}. With the existence of new data from the
NA61/SHINE Collaboration which cover now the inclusive spectra of
$\pi^{\pm}$, $K^{\pm}$, $p$ and $\bar p$ produced in inelastic $p+p$
interactions at all the available SPS energies~\cite{NA61_pp}, we
consider it necessary and important to present in this paper the
first comparison of these most recent data to the UrQMD simulations.
This we do for all identified particles at wide energy range from 20
to 158~GeV.

%We note that an earlier analysis has been performed in
%Ref.~\cite{Vovchenko}, studying the precedent NA61/SHINE dataset
%including only spectra of negatively charged pions produced in
%inelastic $p+p$ collisions~\cite{NA61_pp_previous}. The importance
%of the extension of this former analysis to positive pions, kaons
%and (anti-)protons will become evident in the course of this paper.
%Also in Ref.~\cite{Uzh} the possible improvements of the UrQMD model
%have been studied for a better description of experimental data on
%$\pi$-meson production in $p+p$ and $p+C$ interactions at SPS
%energies, such as accounting of $\eta$-meson decays and inclusion of
%the low mass diffraction dissociation.

The paper is organized as follows. In Sec.~\ref{UrQMD} we provide a
brief description of the UrQMD transport model. In
Sec.~\ref{Results} we comment on our UrQMD studies of mean particle
multiplicities and rapidity distributions included in
Ref.~\cite{NA61_pp}. Then we present our results for the transverse
momentum spectra and the inverse slope parameters of $\pi^{\pm}$,
$K^{\pm}$, $p$ and $\bar p$ produced in inelastic $p+p$ interactions
at different energies, extracted from the UrQMD simulations, and
compare them to the NA61/SHINE experimental data. Finally,
Sec.~\ref{Summary} is dedicated to the summary and conclusions.

\section{The UrQMD transport model}
\label{UrQMD} The UrQMD (Ultra-relativistic Quantum Molecular
Dynamics) transport model~\cite{UrQMD_1,UrQMD_2} is the
non-equilibrium approach based on an effective solution of the
relativistic Boltzmann equation
\begin{equation}
p^{\mu}\partial_{\mu}f_i(x^{\nu},p^{\nu})=C_i,
\end{equation}
which is used to describe the time evolution of the distribution
functions for particle species $i$ and includes the full collision
term on the right hand side. The underlying degrees of freedom are
hadrons and strings. UrQMD includes $55$ baryon and $32$ meson
species, ground state particle, and all resonance with masses up to
$2.25$~GeV. Full particle-antiparticle, isospin and flavor $SU(3)$
symmetries are applied.

The hadrons propagate on straight lines until the covariant relative
distance between two particles gets smaller than a critical distance
given by the corresponding total cross section. The elementary cross
sections are calculated by the detailed balance or the additive
quark model or fitted and parametrized according to the available
experimental data. For resonance excitations and decays the
Breit-Wigner formalism is used.

The initial high energy phase of the reaction is modeled via the
excitation and fragmentation of strings treated according to the
LUND model~\cite{LUND_1,LUND_2,LUND_3}; for hard collisions with
large momentum transfer ($Q>1.5$~GeV) PYTHIA~\cite{pythia} is used.

In the present study, we use the most recent version of UrQMD
transport model, UrQMD v3.4~\cite{UrQMD_3}, which has been
successfully applied to describe particle yields and transverse
dynamics in the energy range from $E_{lab}=2$ to
$160$~AGeV~\cite{UrQMD_test_1}.

\section{Results}
\label{Results} In this Section we present the results for the
transverse momentum spectra and the inverse slope parameters of
$\pi^{\pm}$, $K^{\pm}$, $p$ and $\bar p$ produced in inelastic $p+p$
interactions at different collision energies, obtained from the
UrQMD calculations. Prior to this we note that the UrQMD results for
the rapidity distributions and mean multiplicities of identified
particles for all energies were shown in Fig.~43 and Fig.~44 of
Ref.~\cite{NA61_pp}, respectively. The conclusions from this first
comparison to the NA61/SHINE experimental data included therein can
be summarized as follows.
\begin{figure*}
%\vspace*{3cm}
\centering \subfigure{
\resizebox{0.47\textwidth}{!}{%
 \includegraphics{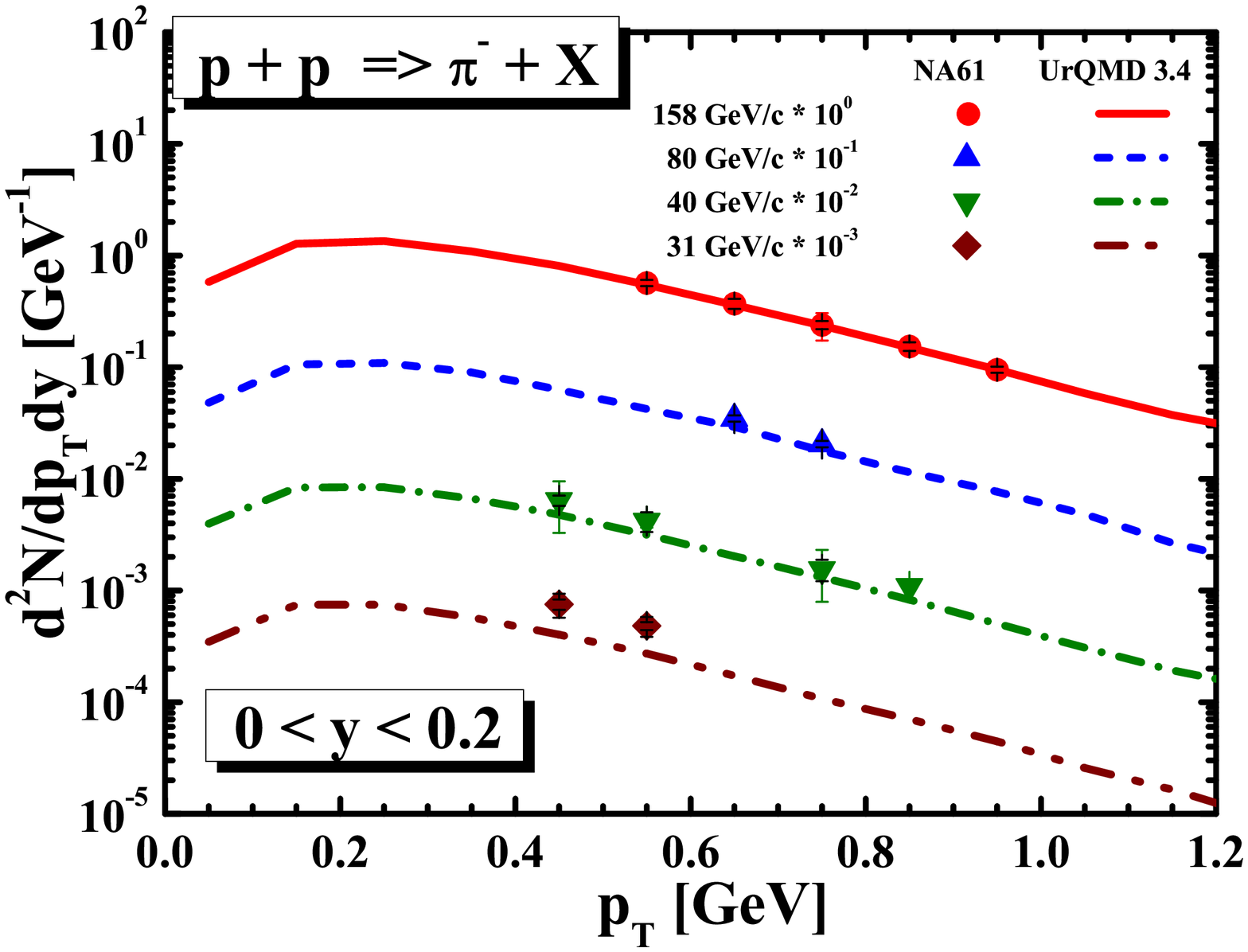}\hspace*{-5cm}
} } \subfigure{
\resizebox{0.47\textwidth}{!}{%
 \includegraphics{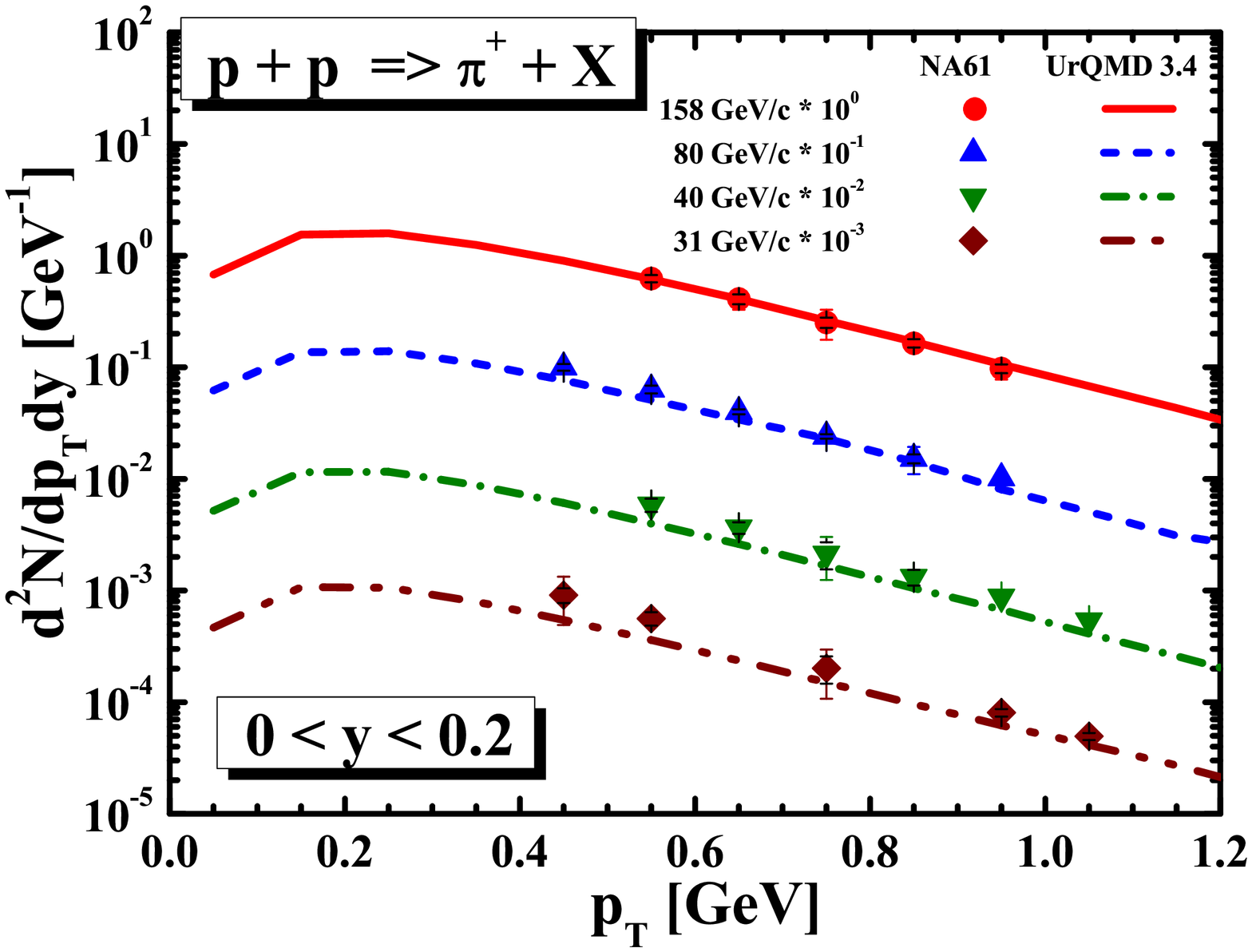}\hspace*{-5cm}
} } \vspace*{-1.7cm}\caption{The UrQMD v3.4 predictions (lines) for
transverse momentum spectra of $\pi^-$ (left) and $\pi^+$ (right)
mesons produced at $0<y<0.2$ in inelastic $p+p$ interactions at beam
momenta of 31, 40, 80 and 158 GeV/$c$, in comparison to experimental
data from the NA61/SHINE Collaboration (symbols).}
\label{dNdptdy_pion_y=0.1}
\end{figure*}
\begin{figure*}
\vspace{0.3cm} \centering \subfigure{
\resizebox{0.47\textwidth}{!}{%
 \includegraphics{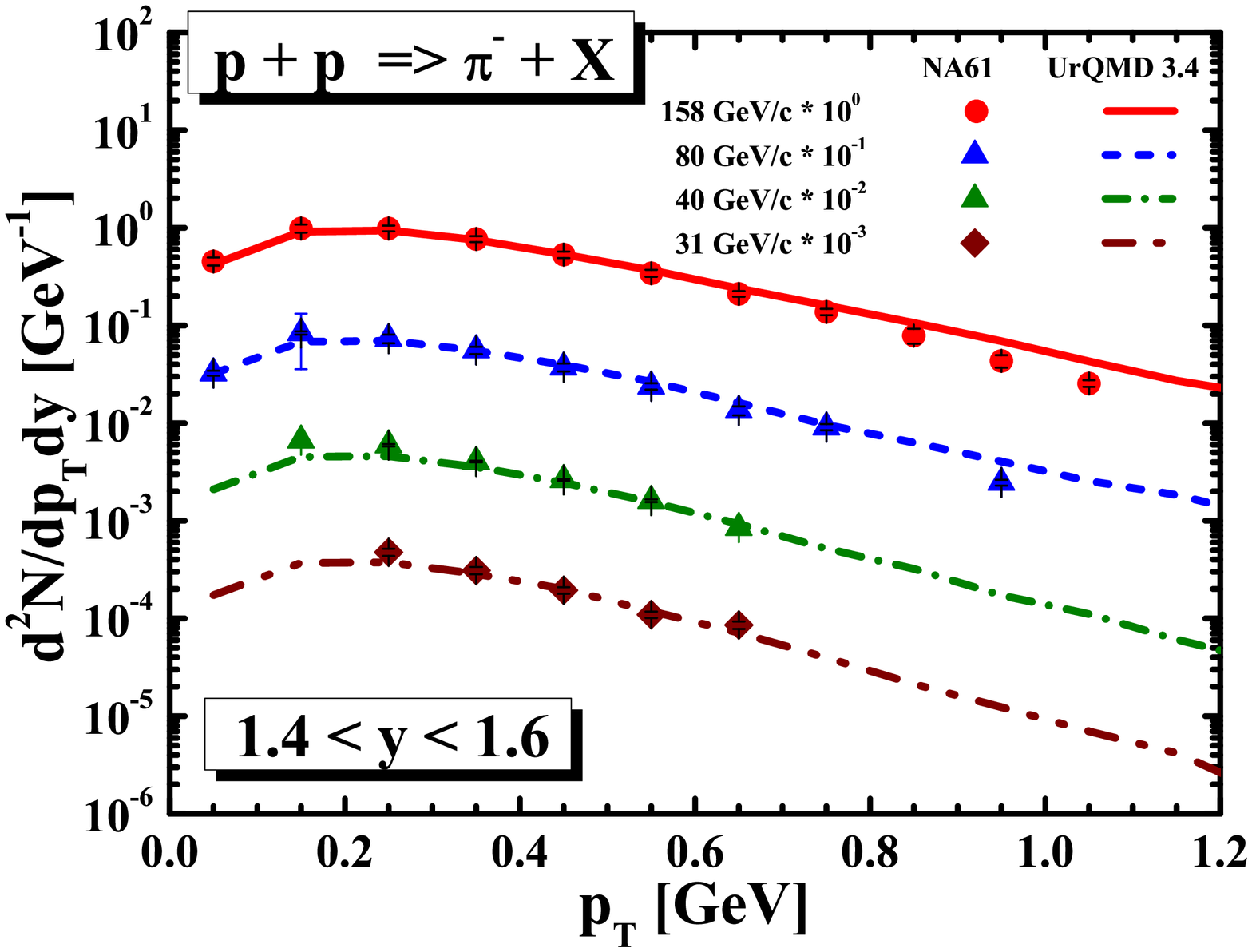}\hspace*{-5cm}
} } \subfigure{
\resizebox{0.47\textwidth}{!}{%
 \includegraphics{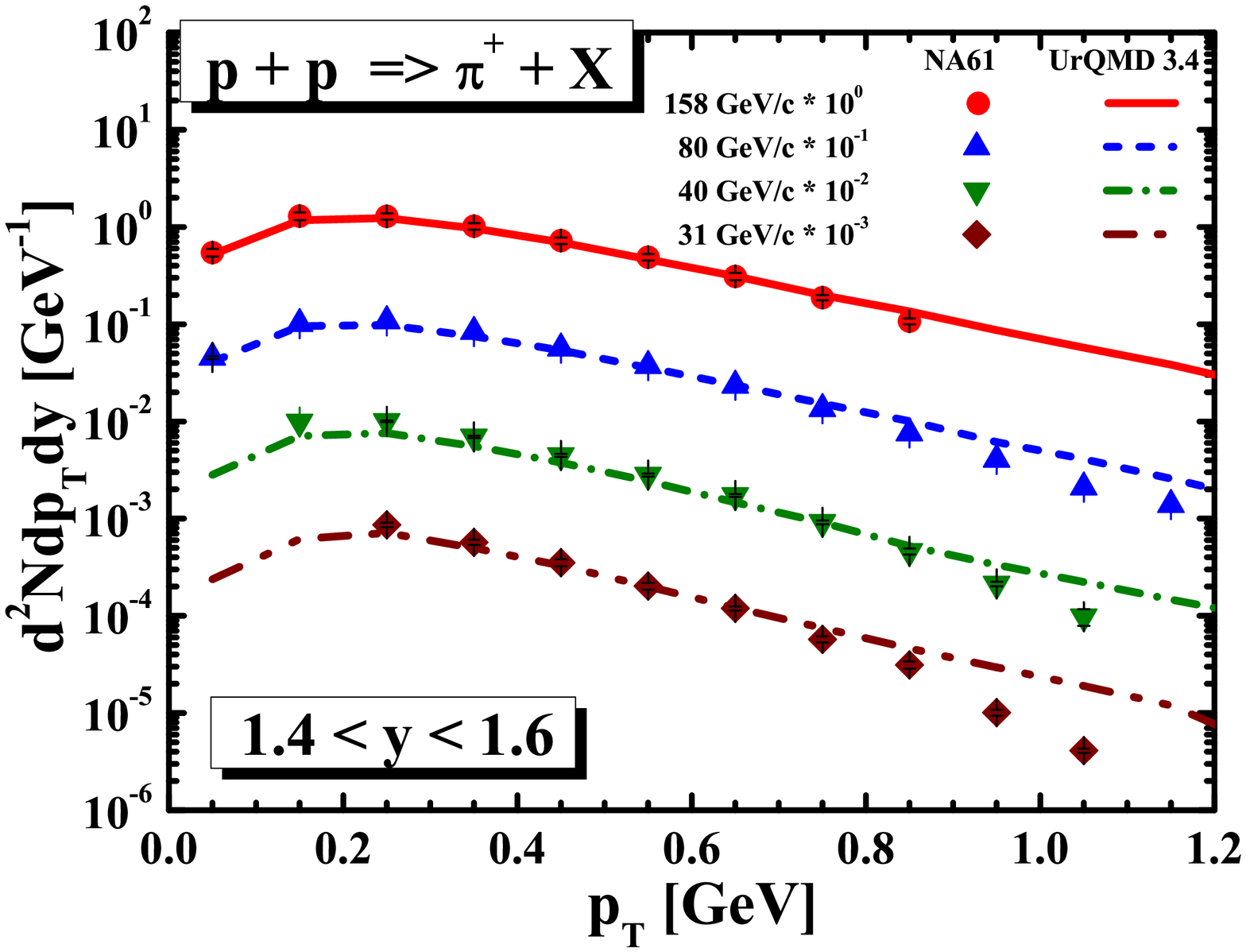}\hspace*{-5cm}
} } \vspace{-1.7cm}\caption{The UrQMD v3.4 predictions (lines) for
transverse momentum spectra of $\pi^-$ (left) and $\pi^+$ (right)
mesons produced at $1.4<y<1.6$ in inelastic $p+p$ interactions at
beam momenta of 31, 40, 80 and 158 GeV/$c$, in comparison to
experimental data from the NA61/SHINE Collaboration (symbols).}
\label{dNdptdy_pion_y=1.5}
%\vspace*{3cm}
\end{figure*}

For {\em pions}, the model describes well the mean $\pi^-$ and
$\pi^+$ multiplicities (see Fig.~44 in Ref.~\cite{NA61_pp} for the
comparison of the mean multiplicities here and below in the text)
and $dN/dy$ spectra (see Fig.~43 in Ref.~\cite{NA61_pp} for the
comparison of $dN/dy$ spectra here and below in the text) at high
SPS energies. It gradually underestimates the mean multiplicities of
both particles at lower energies. In the 20-31 GeV/c beam momentum
regime, the corresponding discrepancies may reach 30\% for the mean
$\pi^-$ and $\pi^+$ multiplicities and a factor of two (up to 30\%)
for negative (positive) pion $dN/dy$ values at $y\approx 0$. We note
that we always address rapidity $y$ in the collision center-of-mass
system.

For {\em kaons}, the UrQMD model does not match the energy
dependence of their mean multiplicity. For $K^-$ mesons, it slightly
underestimates their mean multiplicities at beam momenta of 20 and
31 GeV/c, underestimates it by about 30\% at 40 GeV/c, and
overestimates it by up to 30\% at higher beam momenta. A roughly
similar behavior is apparent for $dN/dy$ distributions. For $K^+$,
the model provides a rough description of mean multiplicities and
$dN/dy$ spectra for the two top beam momenta, but systematically
underestimates both observables at lower energies, with
discrepancies reaching 40\% of the mean multiplicity and a factor of
two for $dN/dy(y\approx 0)$, respectively.

\begin{figure*}
%\vspace*{3cm}
\centering \subfigure{
\resizebox{0.47\textwidth}{!}{%
 \includegraphics{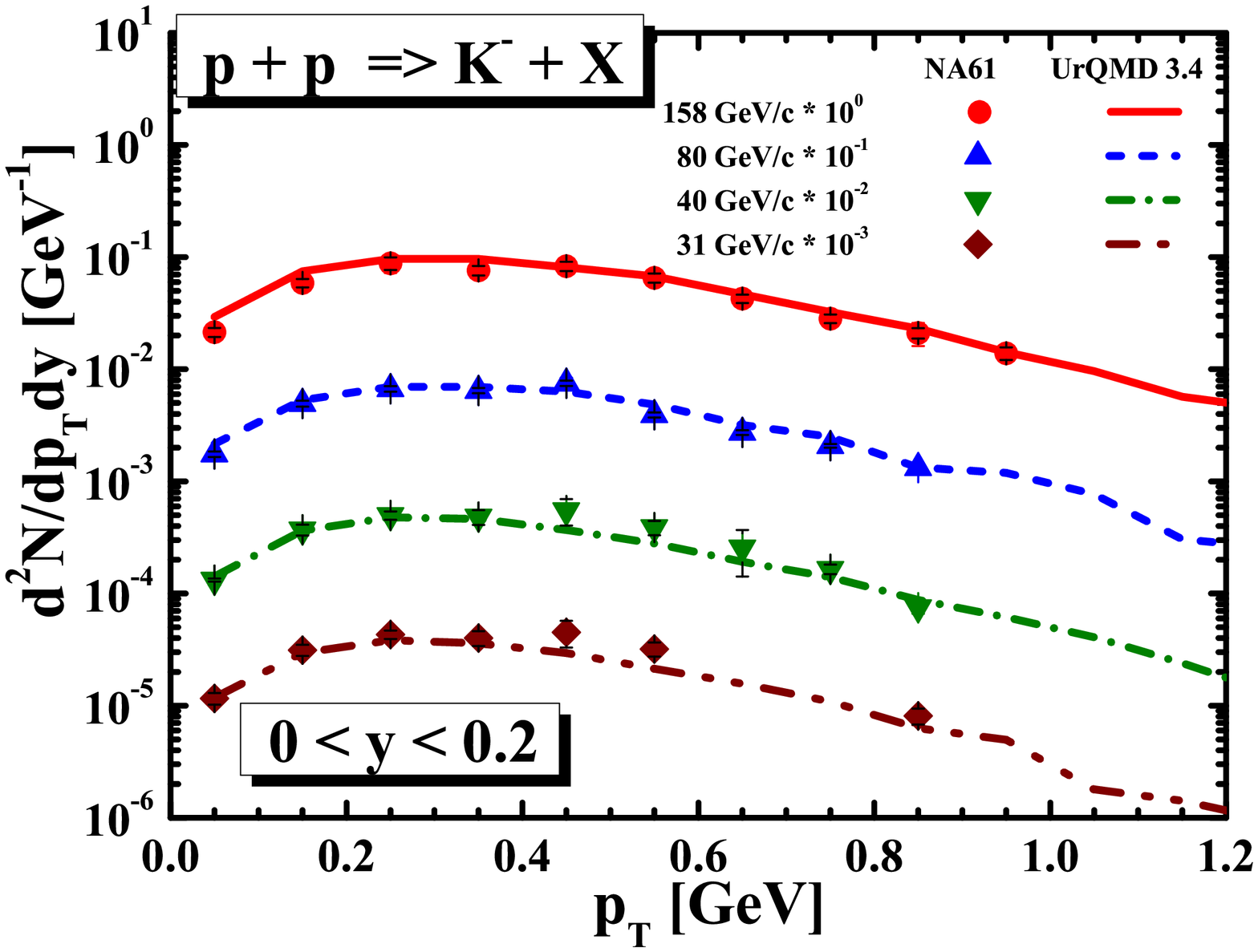}\hspace*{-5cm}
} } \subfigure{
\resizebox{0.47\textwidth}{!}{%
 \includegraphics{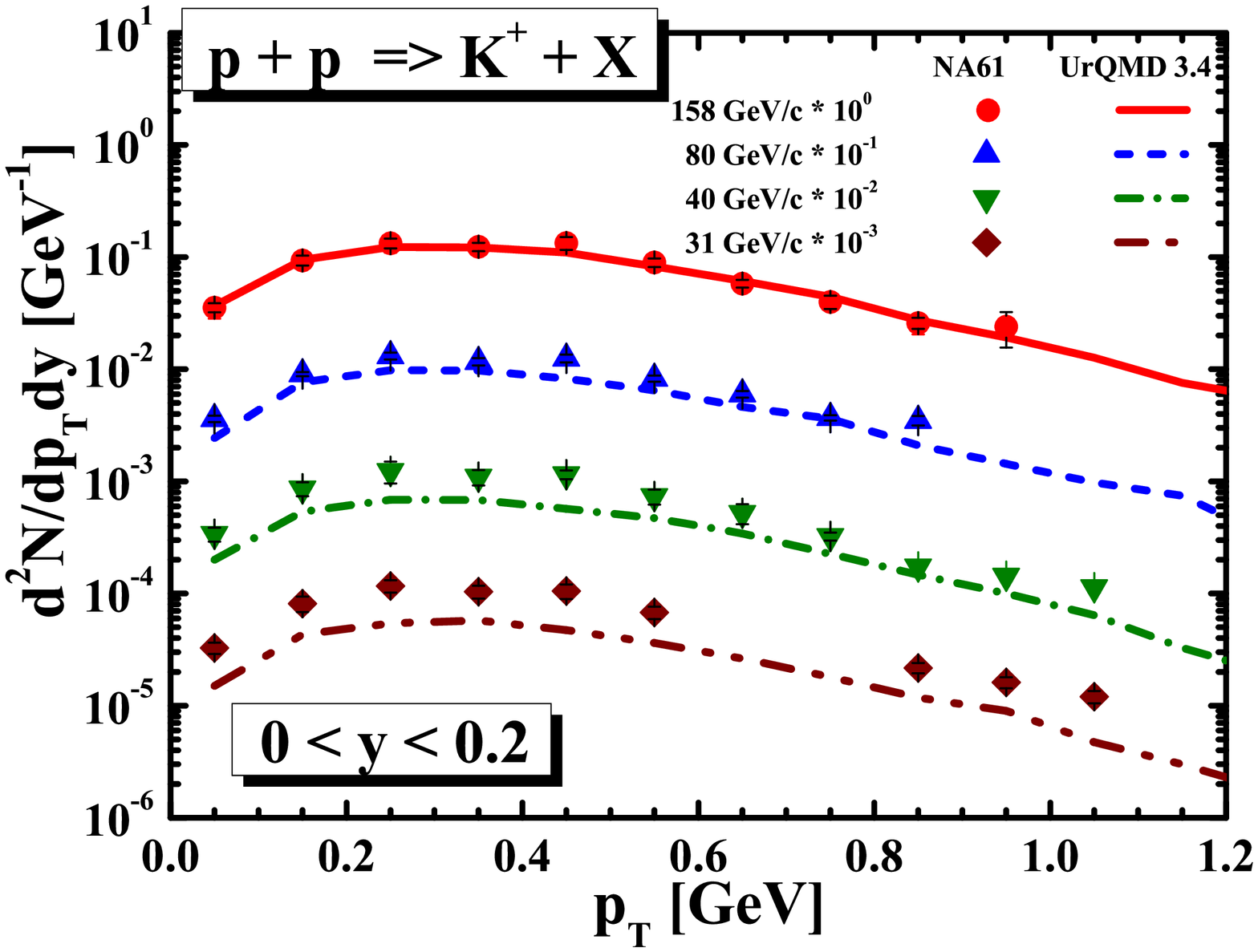}\hspace*{-5cm}
} } \vspace{-1.7cm}\caption{The UrQMD v3.4 predictions (lines) for
transverse momentum spectra of $K^-$ (left) and of $K^+$ (right)
produced at $0<y<0.2$ in inelastic $p+p$ interactions at 31, 40, 80
and 158 GeV/$c$, in comparison to experimental data from the
NA61/SHINE Collaboration (symbols). The small cusps on the lines
correspond to statistical fluctuations in our model calculation.}
\label{dNdptdy_Kaon_y=0.1}
%Monte Carlo simulation.} \label{dNdptdy_Kaon_y=0.1}
\end{figure*}
\begin{figure*}
\vspace{0.3cm} \centering \subfigure{
\resizebox{0.47\textwidth}{!}{%
 \includegraphics{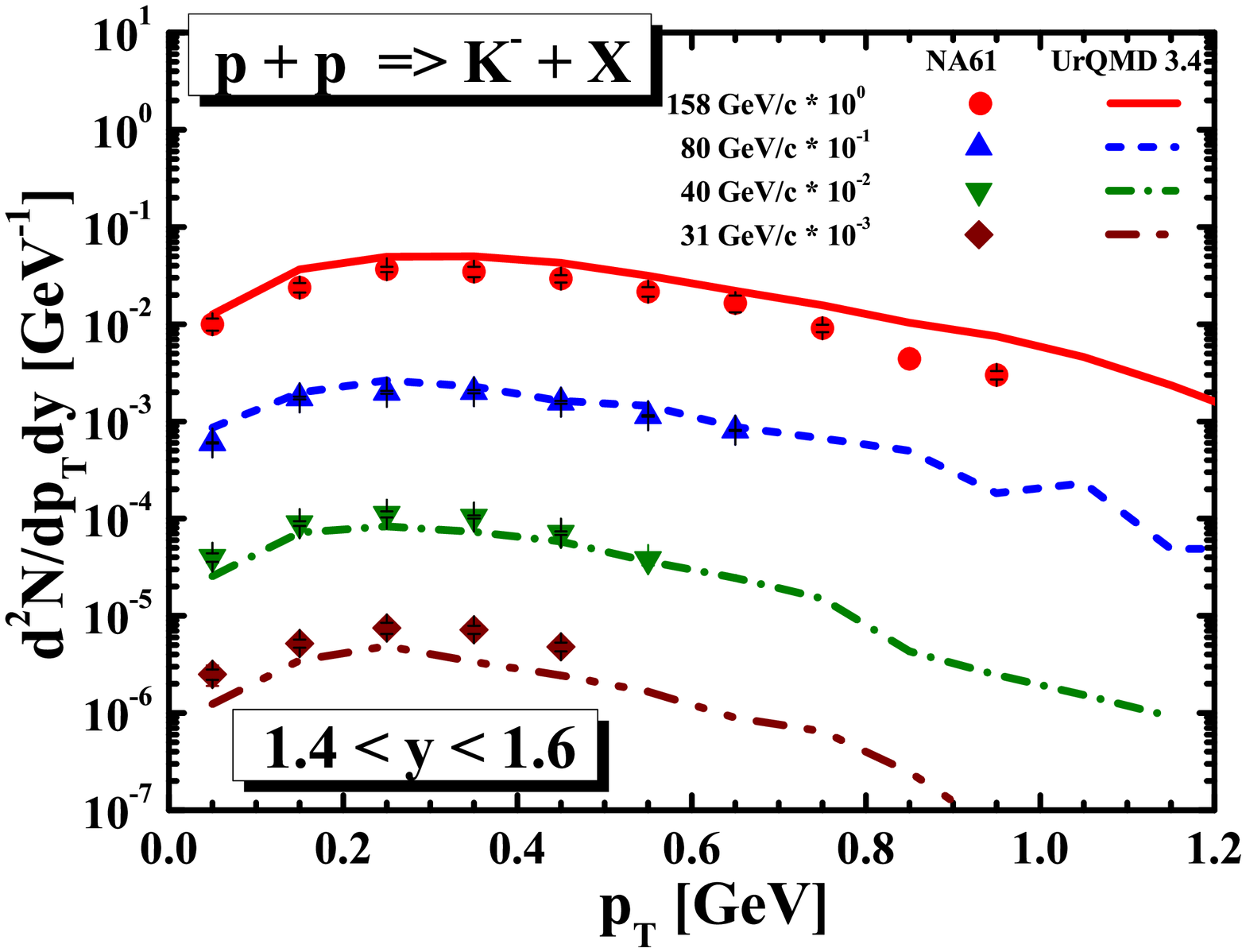}\hspace*{-5cm}
} } \subfigure{
\resizebox{0.47\textwidth}{!}{%
 \includegraphics{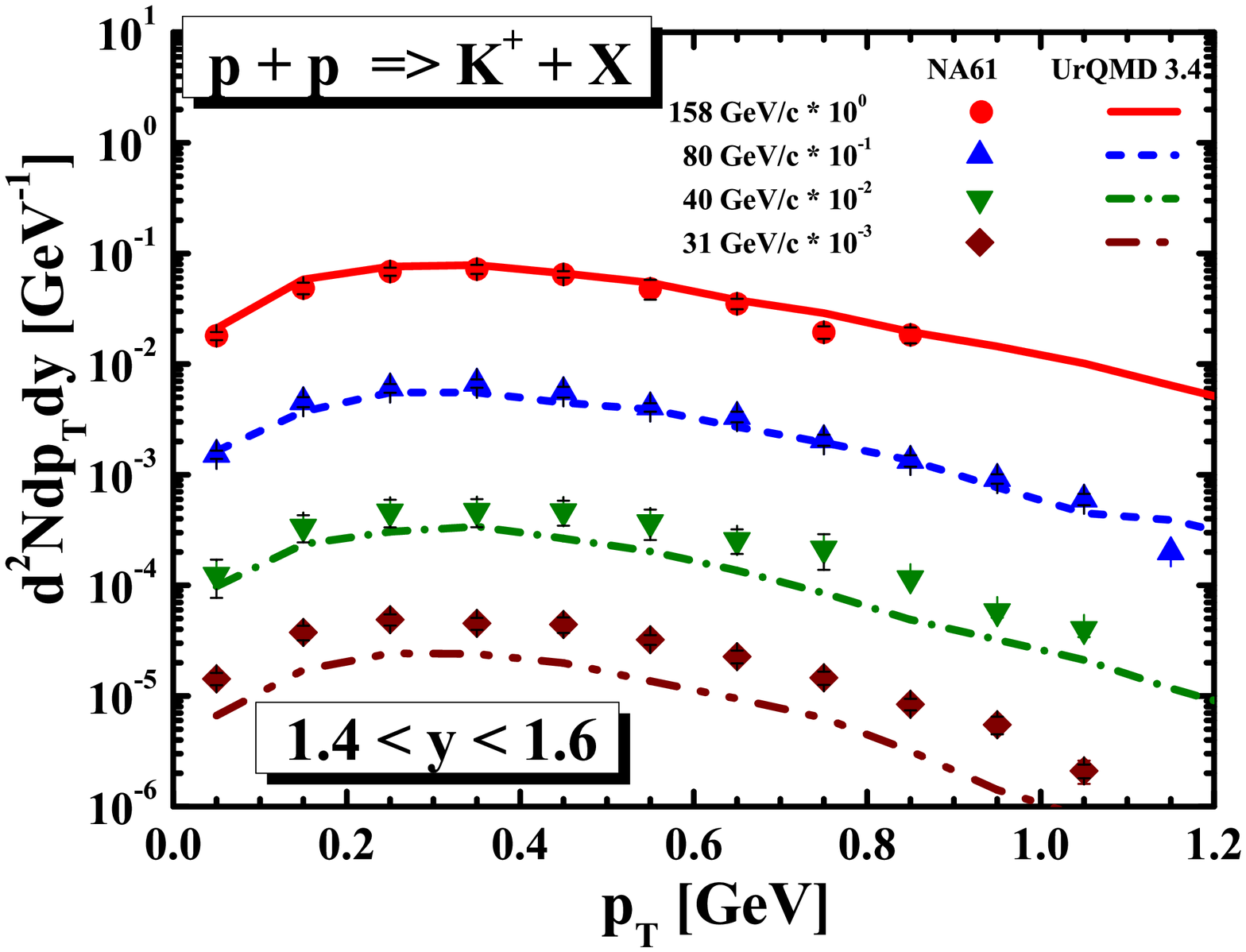}\hspace*{-5cm}
} } \vspace{-1.7cm}\caption{The UrQMD v3.4 predictions (lines) for
transverse momentum spectra of $K^-$ (left) and of $K^+$ (right)
produced at $1.4<y<1.6$ in inelastic $p+p$ interactions at 31, 40,
80 and 158 GeV/$c$, in comparison to experimental data from the
NA61/SHINE Collaboration (symbols). The small cusps on the lines
correspond to statistical fluctuations in our model calculation.}
\label{dNdptdy_Kaon_y=1.5}
%Monte Carlo simulation.} \label{dNdptdy_Kaon_y=1.5}
%\vspace*{3cm}
\end{figure*}
\begin{figure*}
%\vspace*{3cm}
\centering \subfigure{
\resizebox{0.47\textwidth}{!}{%
 \includegraphics{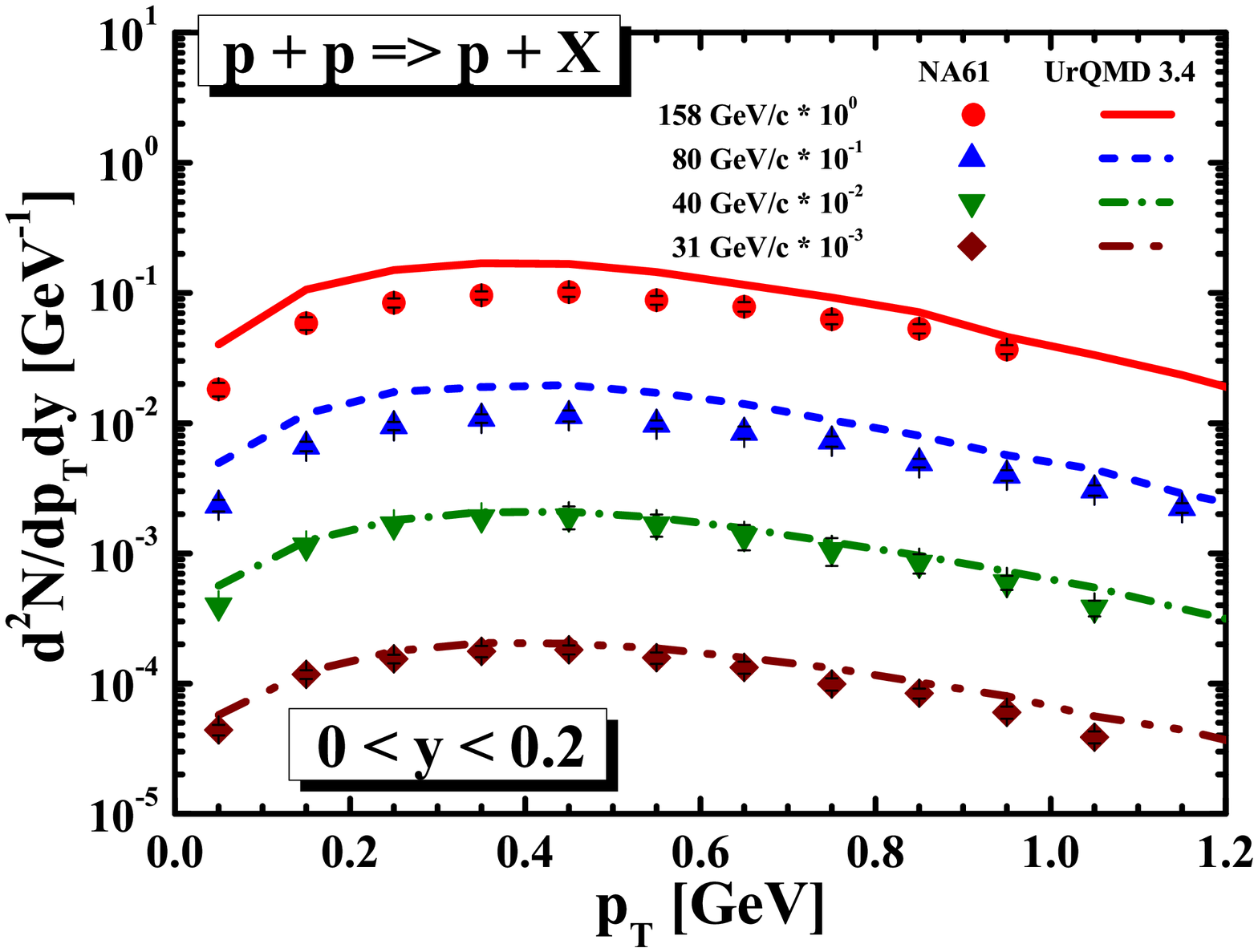}\hspace*{-5cm}
} } \subfigure{
\resizebox{0.47\textwidth}{!}{%
 \includegraphics{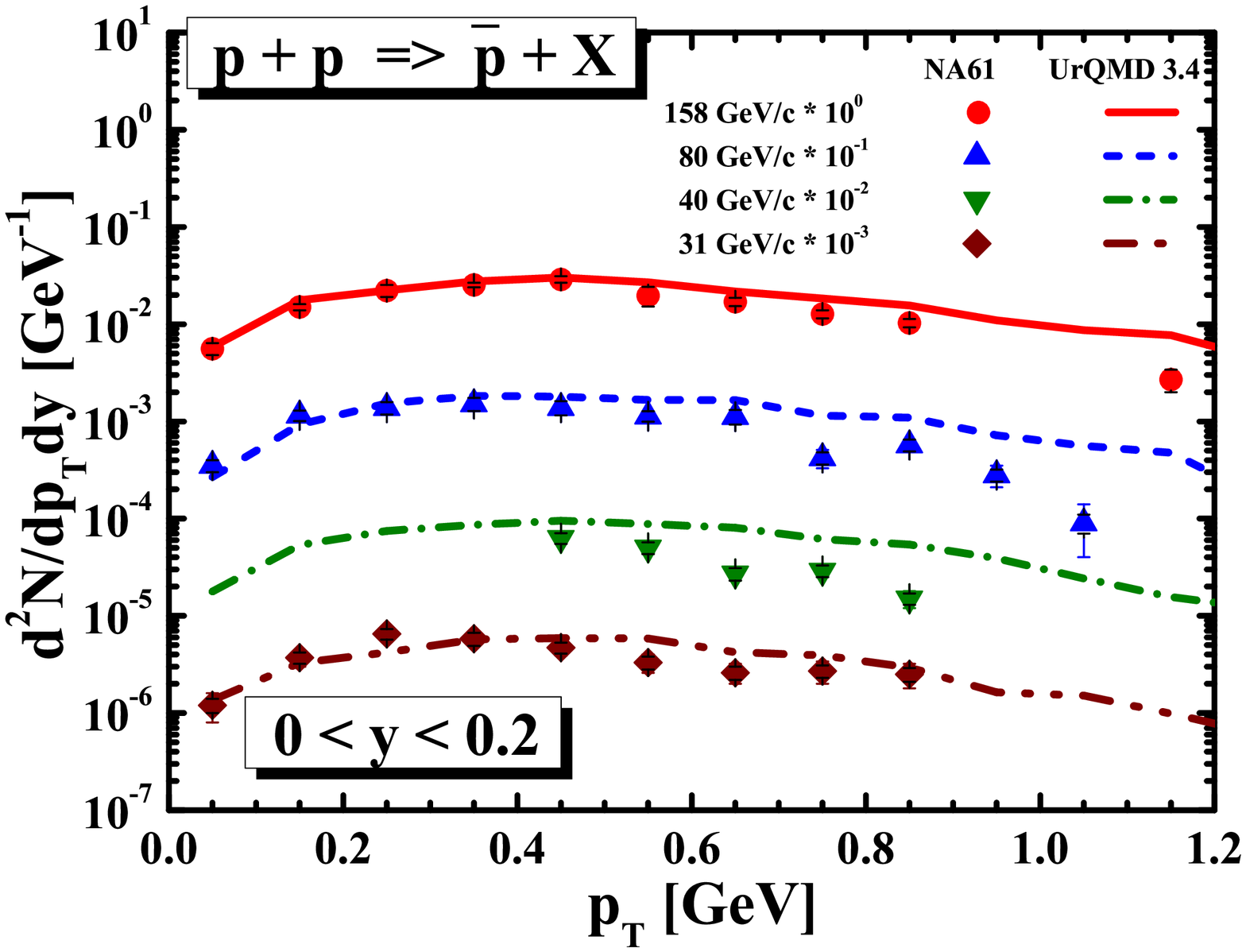}\hspace*{-5cm}
} } \vspace{-1.7cm}\caption{The UrQMD v3.4 predictions (lines) for
transverse momentum spectra of protons (left) and antiprotons
(right) produced at $0<y<0.2$ in inelastic $p+p$ interactions at 31,
40, 80 and 158 GeV/$c$, in comparison to experimental data from the
NA61/SHINE Collaboration (symbols). The small cusps on the lines
correspond to statistical fluctuations in our model calculation.}
\label{dNdptdy_proton_y=0.1}
\end{figure*}

\begin{figure*}
\vspace{0.3cm} \centering \subfigure{
\resizebox{0.47\textwidth}{!}{%
 \includegraphics{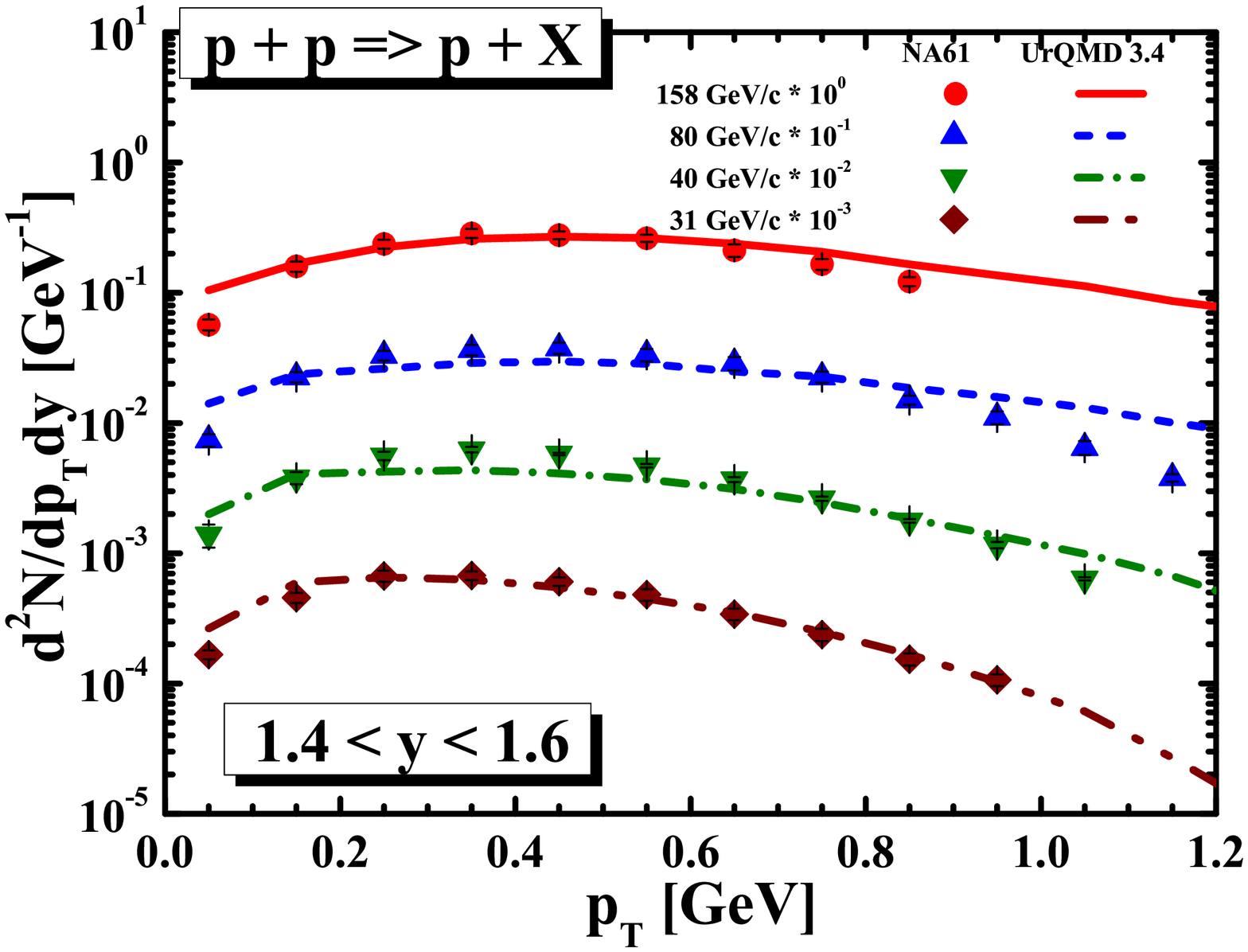}\hspace*{-5cm}
} } \subfigure{
\resizebox{0.47\textwidth}{!}{%
 \includegraphics{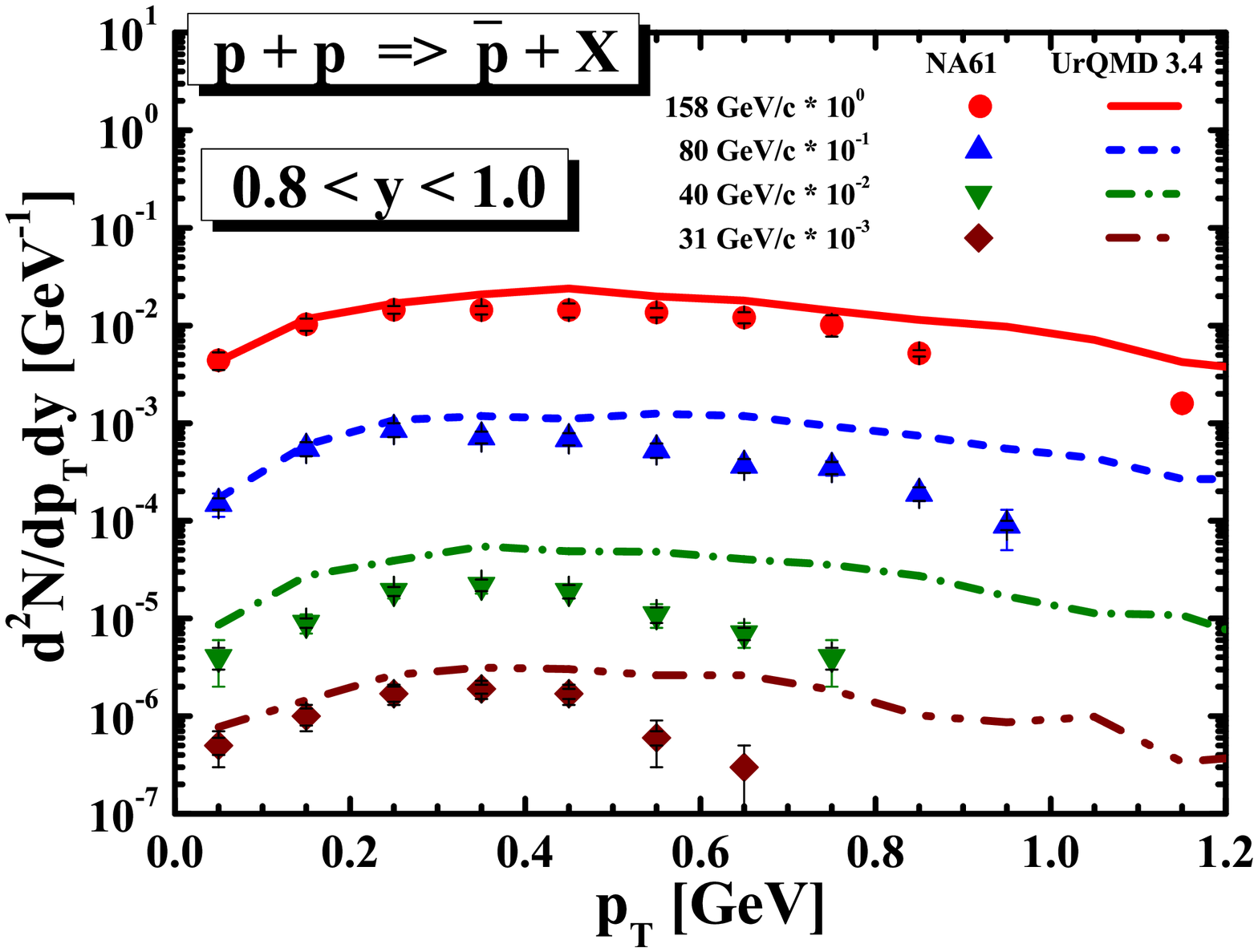}\hspace*{-5cm}
} } \vspace{-1.7cm}\caption{The UrQMD v3.4 predictions (lines) for
transverse momentum spectra of protons (left) and antiprotons
(right) produced at $1.4<y<1.6$ (for protons) and at $0.8<y<1$ (for
antiprotons) in inelastic $p+p$ interactions at 31, 40, 80 and 158
GeV/$c$, in comparison to experimental data from the NA61/SHINE
Collaboration (symbols). The small cusps on the lines correspond to
statistical fluctuations in our model calculation.}
\label{dNdptdy_proton_y=1.5}
\end{figure*}

For {\em protons}, the model matches reasonably well the
experimental $dN/dy$ spectrum for 20 GeV/c beam momentum, with the
exception of the highest rapidity region. For higher beam momenta
the model systematically overpredicts the data in an increasing
range of rapidity. The discrepancy reaches almost a factor of two
for $dN/dy(y\approx 0)$ at 158 GeV/c. For {\em antiprotons}, the
model overpredicts the mean multiplicities by factors from 1.5 to
2.2 for all beam momenta (no data is available at 20 GeV/c).
Consistently, it also overpredicts the $dN/dy$ distributions.

\subsection{Transverse momentum spectra}
\label{transverse} In Figs~\ref{dNdptdy_pion_y=0.1}
and~\ref{dNdptdy_pion_y=1.5} we show the transverse momentum
distributions of $\pi^-$ mesons (left) and $\pi^+$ mesons (right)
produced at central ($0<y<0.2$) and forward rapidity ($1.4<y<1.6$)
in inelastic $p+p$ collisions at 31, 40, 80 and 158 GeV/$c$. We note
that the projectile beam rapidity is respectively 2.094, 2.223,
2.569 and 2.909 for the corresponding beam
momenta~\cite{NA61_pp_previous}. Our model predictions are compared
to the experimental data from the NA61/SHINE
Collaboration~\cite{NA61_pp} obtained at the same pion rapidity. The
symbols with error bars correspond to the experimental data and the
lines to our UrQMD simulations. For better visibility, the
experimental and model spectra are scaled by common factors at the
different beam momenta.

One observes that at central rapidity the UrQMD model describes well
the transverse momentum spectra of both positive and negative pions
at 158~GeV/$c$ , but underestimates their yields at the lower beam
energies and for the whole transverse momentum range where the
experimental points are available. On the other hand, at forward
rapidity the model gives a good description of the transverse
momentum distribution of pions for $p_T<0.8$~GeV/$c$, but
overestimates the pion yield for $p_T>0.8$~GeV/$c$  at all the
considered beam energies (whenever there is available experimental
data). Thus a specific pattern of deviations of the two-dimensional
$d^2N/dp_Tdy(y,p_T)$ distribution emerges between data and model,
where the pion midrapidity density and high $p_T$ pion yield at
forward rapidity go respectively below and above the experimental
data with decreasing collision energy. This results, after
integration over transverse momentum and rapidity, in the
discrepancy between the mean pion multiplicities and $dN/dy$
distributions in the experimental data and the UrQMD model as a
function of decreasing collision energy, which we discussed above.

The transverse momentum spectra of $K^-$ (left) and $K^+$ (right)
mesons at central and at forward rapidity are shown in
Figs~\ref{dNdptdy_Kaon_y=0.1} and~\ref{dNdptdy_Kaon_y=1.5},
respectively. The model calculations overpredict the $K^-$
experimental data at top collision energy at forward rapidity and
$p_T$. At low collision energies (31 and 40~GeV/$c$ beam momentum),
UrQMD predicts smaller yields of kaons than visible in the
experimental data; this is generally valid for all presented values
of kaon rapidity and transverse momentum and will result in the
underprediction of mean $K^-$ multiplicities which we addressed
above. For $K^+$ mesons a fair agreement between data and model is
achieved at 80 and 158 GeV beam energy, but at lower beam momenta
the UrQMD calculations visibly underestimate the $K^+$ yield as a
function of both rapidity and $p_T$. This is consistent with what
has been said above on the mean $K^+$ multiplicities and rapidity
distributions, but emphasizes the lack of quantitative description
of positive kaon production in the lower SPS energy regime by the
UrQMD model. This remains important for the interpretation of the
energy dependence of strangeness production, claimed to display some
similarity between $p+p$ and central Pb+Pb collisions as discussed
in Ref.~\cite{Aduszkiewicz-qm2017}.

Finally, in Figs~\ref{dNdptdy_proton_y=0.1}
and~\ref{dNdptdy_proton_y=1.5} we show the transverse momentum
distributions of protons (left) and antiprotons (right) produced at
central and at forward rapidity in inelastic $p+p$ collisions at 31,
40, 80 and 158 GeV/$c$. By ``forward rapidity'' we mean the usual
$1.4<y<1.6$ for protons but $0.8<y<1$ for antiprotons. The UrQMD
calculation systematically overpredicts the central rapidity proton
yield in the whole considered range of transverse momentum, an
effect which is much less visible, or absent, at forward rapidity.
This difference between central and forward rapidity is most evident
at top SPS energies, indicating quantitative differences in
transport of baryon number down to low values of rapidity in
physical $p+p$ events and in the UrQMD code. Here we note that the
earlier, high quality proton $dN/dy$ data at 158 GeV/c from the NA49
Collaboration~\cite{na49-pp-proton} are in good agreement with the
experimental result from NA61/SHINE as discussed in
Ref.~\cite{NA61_pp}.

For antiprotons, in spite of the evidently larger relative
uncertainties in the experimental data, it is clear that the UrQMD
model systematically overestimates the NA61/SHINE result. The
discrepancy tends to be larger for forward rapidity and at lower
collision energies, where also the shape of the transverse momentum
distribution differs significantly between the data and the model.

\subsection{The inverse slope parameter}
\label{inverse_slope}

\begin{figure*}
\centering \subfigure{
\resizebox{0.47\textwidth}{!}{%
 \includegraphics{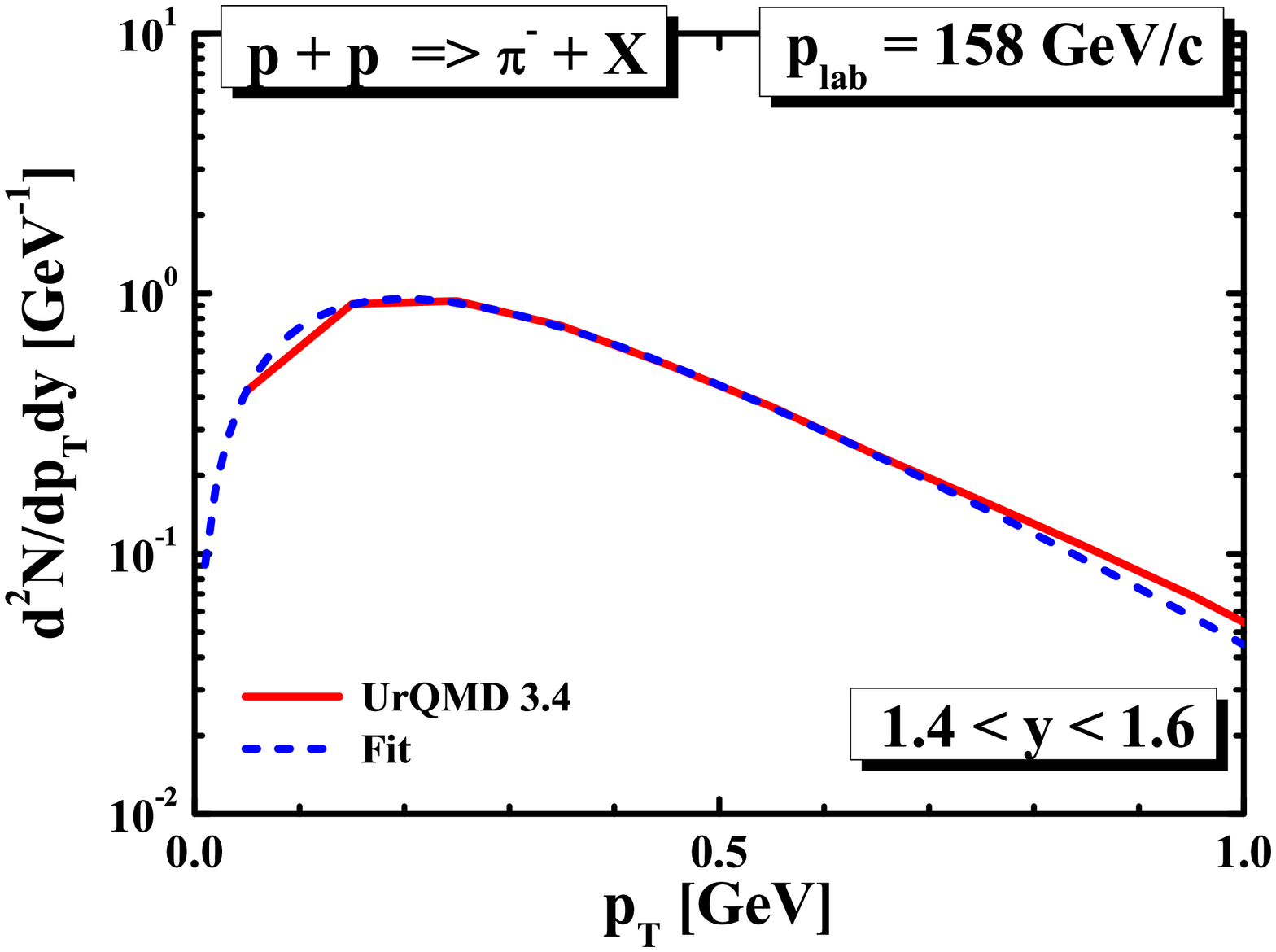}\hspace*{-5cm}
} } \subfigure{
\resizebox{0.47\textwidth}{!}{%
 \includegraphics{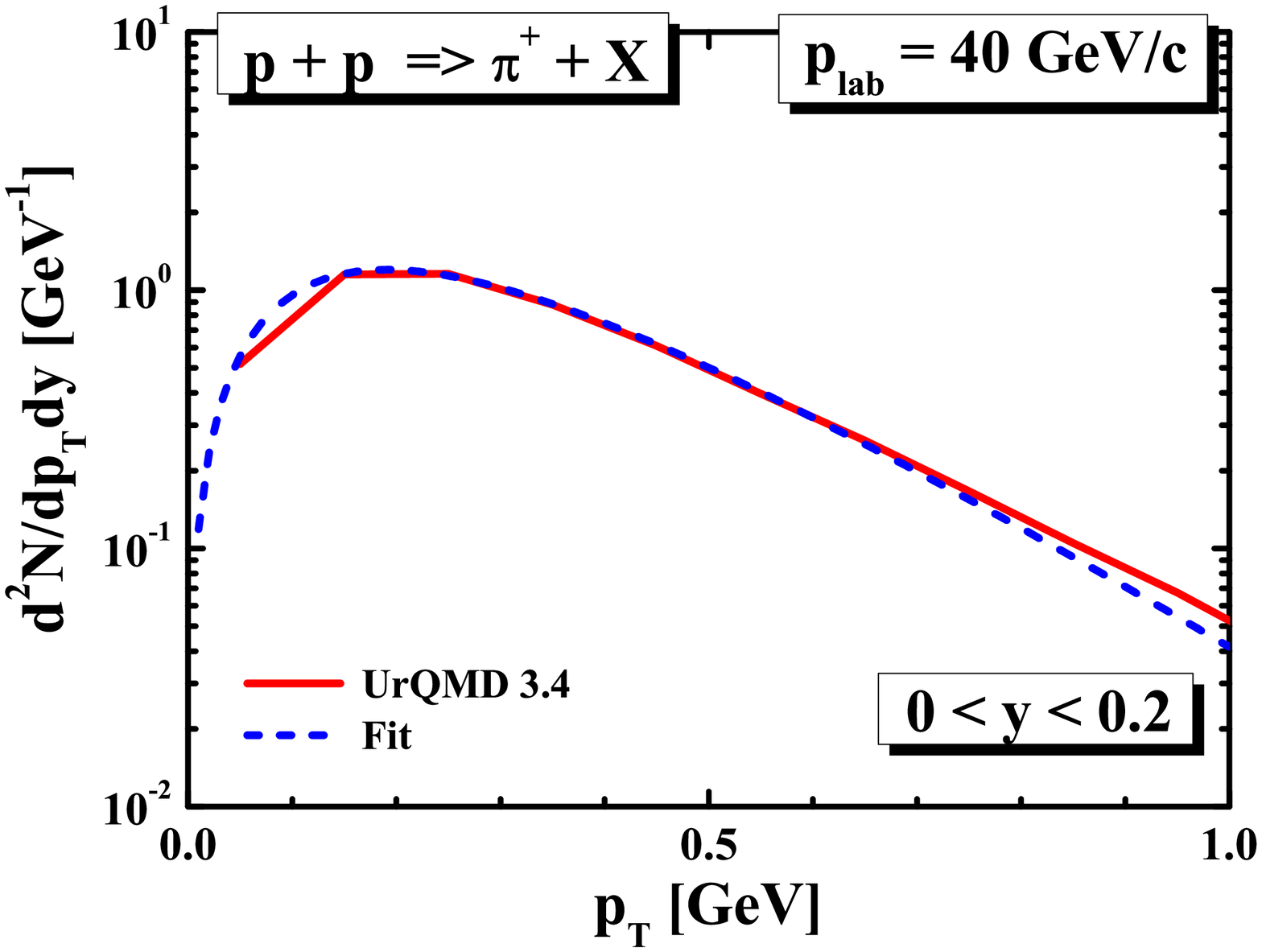}\hspace*{-5cm}
} } \vspace{-1.5cm}\caption{The UrQMD v3.4 simulations (solid lines)
of transverse momentum spectra of $\pi^-$ mesons produced in
inelastic $p+p$ interactions at $1.4<y<1.6$, at 158~GeV/$c$ beam
momentum (left), and of $\pi^+$ mesons produced at $0<y<0.2$, at
40~GeV/$c$ beam momentum (right). The dashed lines represent the
corresponding fits, made according to Eq.~(\ref{fit_function}).}
%\vspace{-0.7cm}
\label{fit_pion}
\end{figure*}
\begin{figure*}
\centering \subfigure{
\resizebox{0.47\textwidth}{!}{%
 \includegraphics{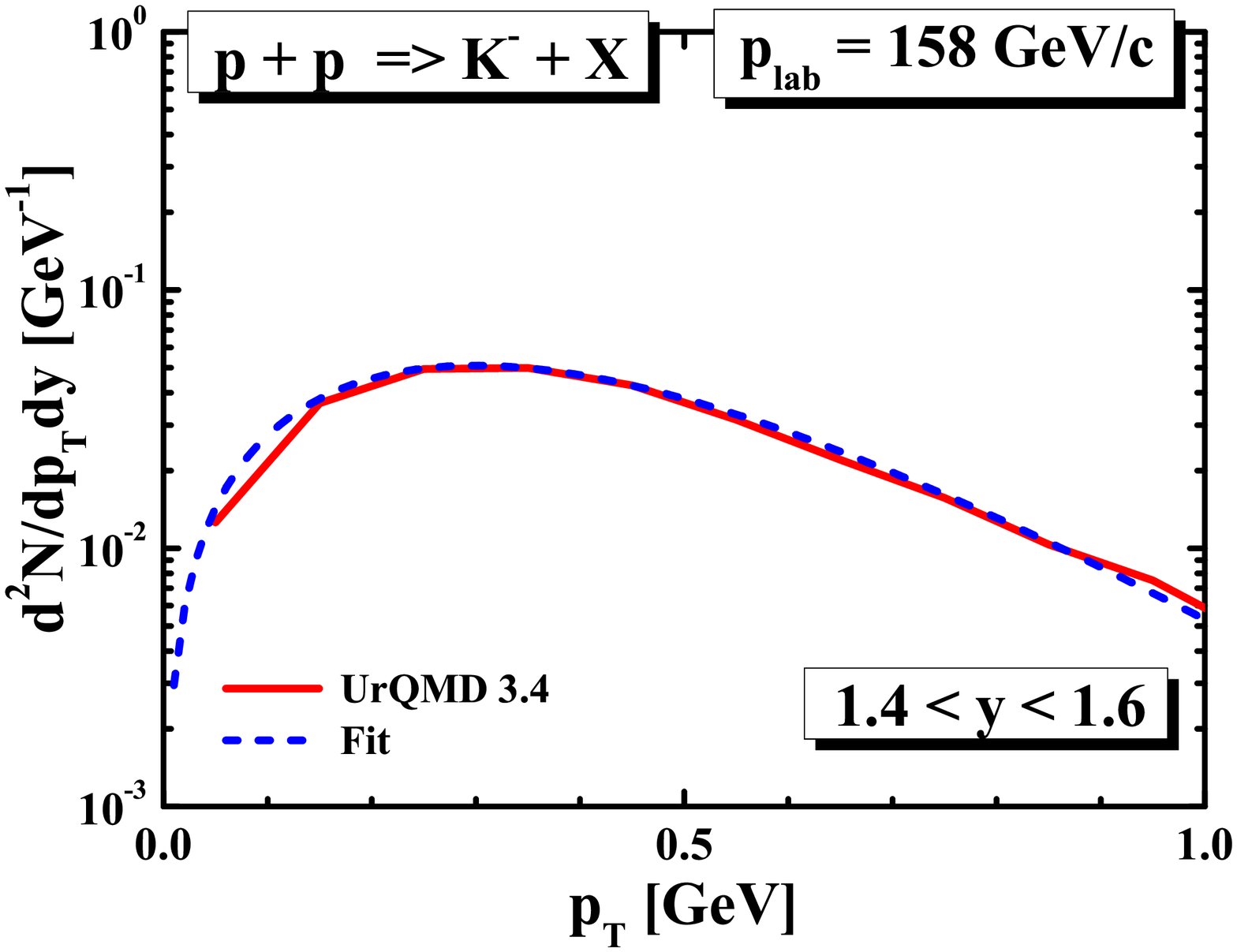}\hspace*{-5cm}
} } \subfigure{
\resizebox{0.47\textwidth}{!}{%
 \includegraphics{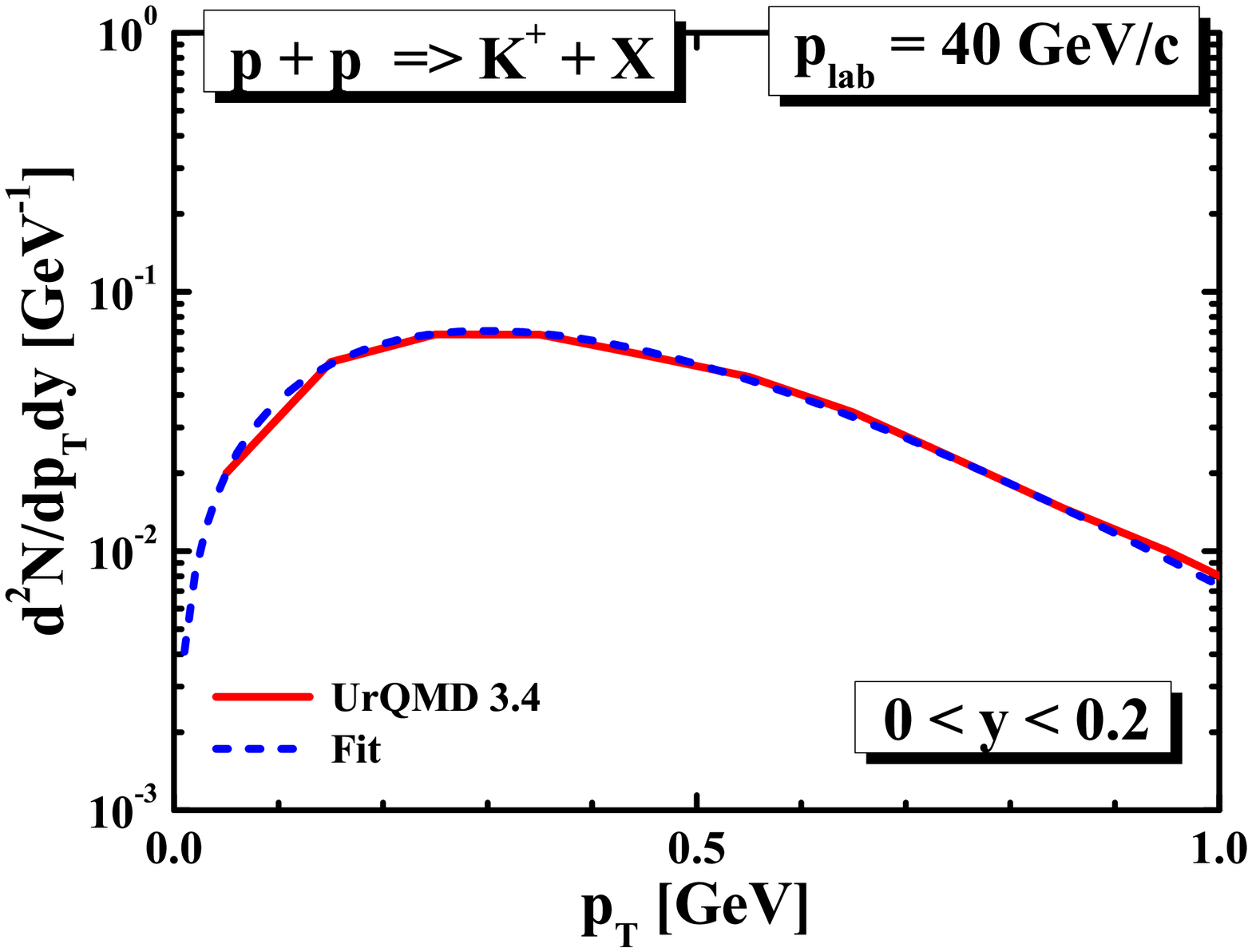}\hspace*{-5cm}
} } \vspace{-1.5cm}\caption{The UrQMD v3.4 simulations (solid lines)
of transverse momentum spectra of $K^-$ mesons produced in inelastic
$p+p$ interactions at $1.4<y<1.6$, at 158~GeV/$c$ beam momentum
(left), and of $K^+$ mesons produced at $0<y<0.2$, at 40~GeV/$c$
beam momentum (right). The dashed lines represent the corresponding
fits, made according to Eq.~(\ref{fit_function}).}
%\vspace{-0.7cm}
\label{fit_kaon}
\end{figure*}

\begin{figure*}
\centering \subfigure{
\resizebox{0.47\textwidth}{!}{%
 \includegraphics{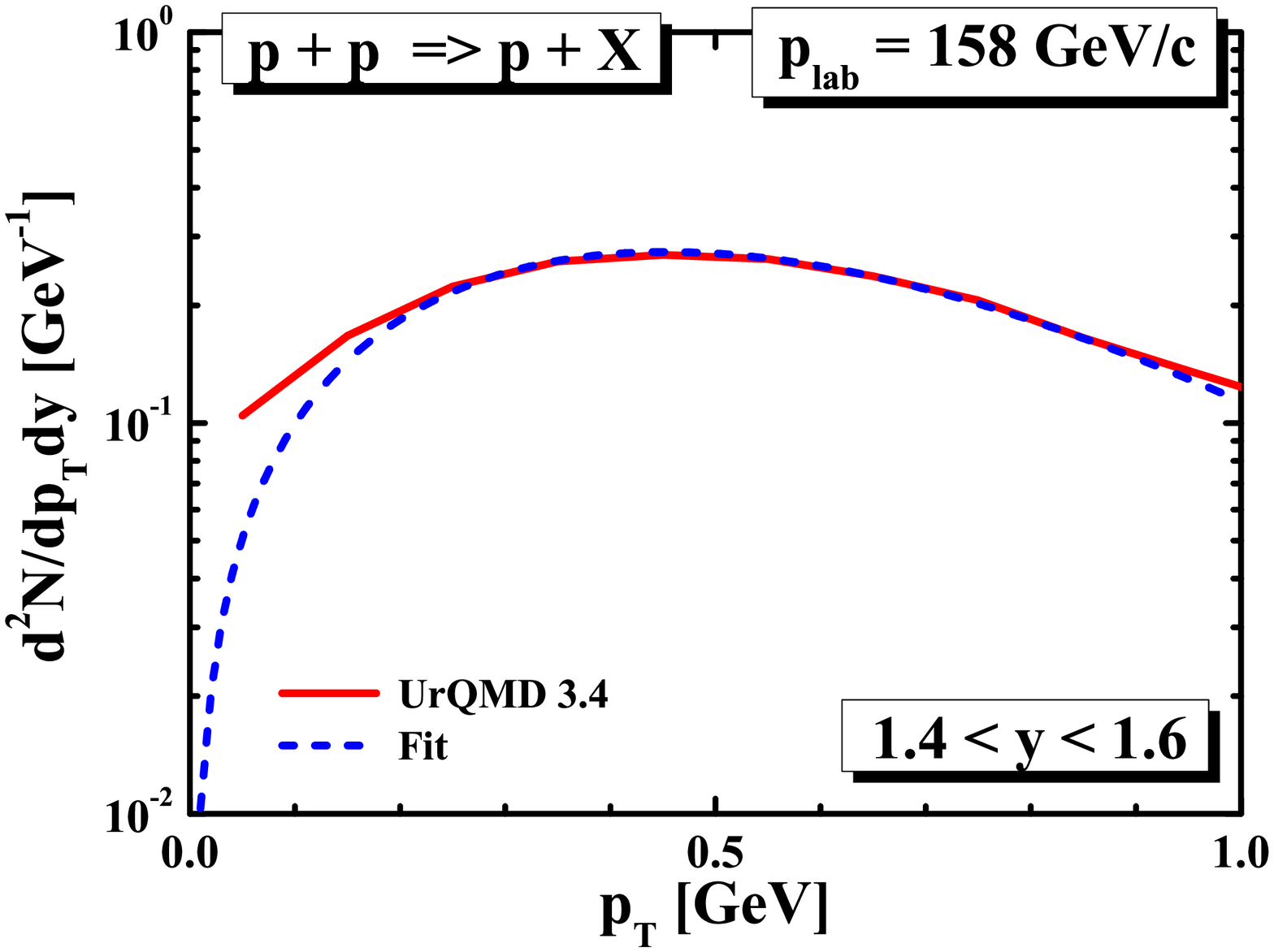}\hspace*{-5cm}
} } \subfigure{
\resizebox{0.47\textwidth}{!}{%
 \includegraphics{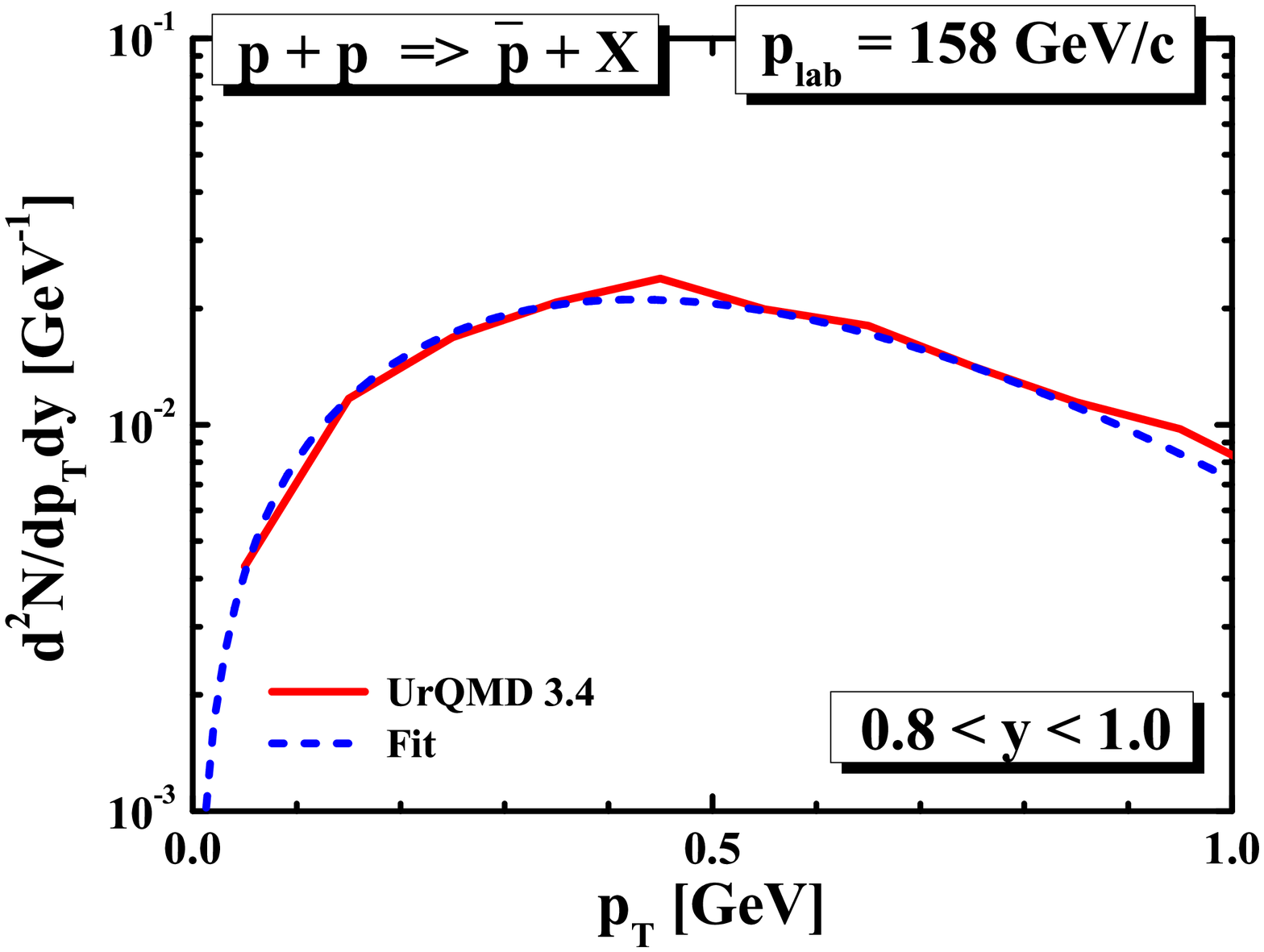}\hspace*{-5cm}
} } \vspace{-1.5cm}\caption{The UrQMD v3.4 simulations (solid lines)
of transverse momentum spectra of protons produced in inelastic
$p+p$ interactions at $1.4<y<1.6$, at 158 GeV/c beam momentum
(left), and of antiprotons produced at $0.8<y<1$ at the same beam
momentum (right). The dashed lines represent the corresponding fits,
made according to Eq.~(\ref{fit_function}).}
%\vspace{-0.7cm}
\label{fit_proton}
\end{figure*}

In order to keep the consistency with the work performed in
Ref.~\cite{NA61_pp}, we attempt to parametrize the results presented
in Figs~\ref{dNdptdy_pion_y=0.1}~-~\ref{dNdptdy_proton_y=1.5} by the
exponential function~\cite{fit_1,fit_2}:
\begin{equation}
\label{fit_function}
\frac{d^2N}{dp_Tdy}=\frac{Sp_T}{T^2+mT}\exp{[-(m_T-m)/T]},
\end{equation}
where $m$ is the mass of the particle, $m_T=\sqrt{m^2+p_T^2}$ is its
transverse mass, $S$ and $T$ are the yield integral and the inverse
slope parameter, respectively. In
Figs~\ref{fit_pion}~-~\ref{fit_proton} we show some examples of this
parametrization fitted to the transverse momentum spectra predicted
by the UrQMD for the different particles produced in inelastic $p+p$
interactions, for a selection of rapidity intervals and colliding
energies. We note that our fits were performed in the whole range of
transverse momentum, while the ranges of the experimental fits were
defined by the available data points and the experimental results
were only shown for those rapidity intervals for which there were
more than six data points in the transverse momentum distribution.
The solid lines represent the UrQMD predictions and the dashed lines
represent the corresponding fits.

One observes that in the mesonic sector, for pions and kaons, the
results of the UrQMD calculations can be reasonably well fitted by
the parametrization from Eq.~(\ref{fit_function}), in the whole
available range of transverse momentum. For pions, some doubt is
induced by the fact that the fit tends to go below the UrQMD
spectrum at $p_T>0.7$~GeV/$c$. We note that a similar behavior, with
respect to the UrQMD predictions, can be seen for some of the
experimental data at least at forward pion rapidity, see
Fig.~\ref{dNdptdy_pion_y=1.5}. Thus the agreement between two
inverse slopes, one fitted to the experimental $p_T$ distribution
and the other fitted to that predicted by the model will not be
equivalent to the agreement between the two distributions
themselves.

For kaons, Fig.~\ref{fit_kaon}, no significant deviations between
the spectrum predicted by the model and the fitted
parametrization~(\ref{fit_function}) are seen. This is reminiscent
of the good agreement between experimental kaon spectra and the same
parametrization~(\ref{fit_function}) which was demonstrated in
Ref.~\cite{NA61_pp}. On the other hand, one should remember the
discrepancies between the experimental data and the UrQMD code shown
in Figs~\ref{dNdptdy_Kaon_y=0.1} and~\ref{dNdptdy_Kaon_y=1.5}.

For protons, Fig.~\ref{fit_proton} (left panel), it is not possible
to fit our results for all the values of $p_T$ simultaneously when
using the parametrization~(\ref{fit_function}). The presented fit
attempt very significantly underpredicts the proton yield at low
transverse momenta. Finally, an attempt to fit the
parametrization~(\ref{fit_function}) to the antiproton $p_T$
spectrum is presented in  Fig.~\ref{fit_proton} (right panel). After
a detailed inspection, we conclude that our statistics does not
allow us to conclusively judge on the fit quality, even at top SPS
energy where the produced $\bar{p}$ yield is the highest.

\subsection{Inverse slope as a function of rapidity}
\label{inverse_slope_rapidity} Similarly to the work presented in
Ref.~\cite{NA61_pp}, we apply the fit of the
parametrization~(\ref{fit_function}) to all the mesons studied in
this paper, at all considered values of rapidity and all colliding
energies. We restrict this part of our study uniquely to mesons in
view of the failure of the fit for protons and its unreliability for
antiprotons as discussed above. The extracted values of the inverse
slope parameter $T$ are shown in Fig.~\ref{inverse_slope} as a
function of meson rapidity at four beam momenta, 31, 40, 80 and
158~GeV/$c$. The solid lines correspond to the UrQMD calculations
and the symbols with the error bars represent the values of $T$
extracted in Ref.~\cite{NA61_pp} from fits to experimental data.

In addition to the UrQMD results, we also show (by dashed lines) the
corresponding fit results obtained from simulations by the EPOS
v1.99 model~\cite{EPOS}; these were published in
Ref.~\cite{NA61_pp}, together with NA61/SHINE data. We mention that
in the EPOS model, the interaction proceeds via excitation of
strings according to the Gribov-Regge theory and then their
fragmentation into hadrons.

\begin{figure*}
\centering
\includegraphics[width=1.25\textwidth]{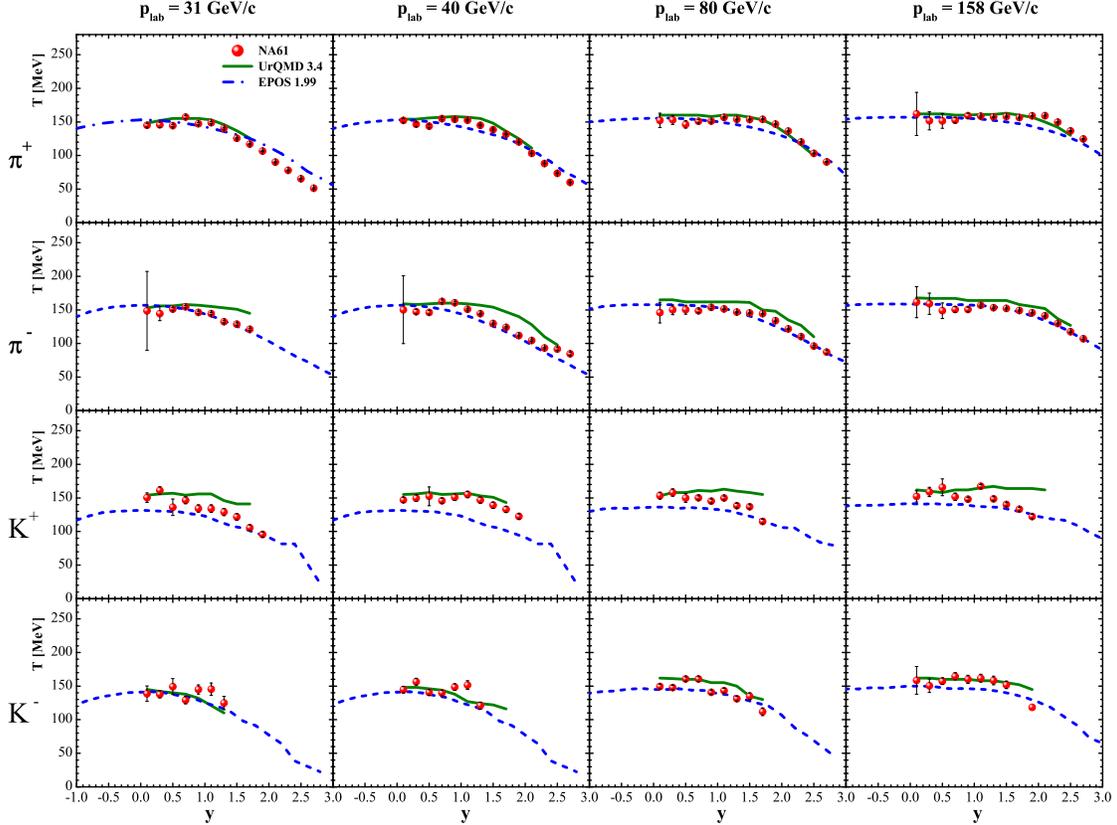}
\vspace*{-3.5cm} \caption{Solid lines: the inverse slope parameter
$T$, extracted from fits with the
parametrization~(\ref{fit_function}) to the transverse momentum
distributions of $\pi^{\pm}$ and $K^{\pm}$ mesons produced in
inelastic $p+p$ interactions at 31, 40, 80 and 158 GeV/$c$, obtained
from UrQMD v3.4 calculations as a function of pion and kaon
rapidity. These results are compared to the values of $T$ obtained
from fits to experimental data from the NA61/SHINE Collaboration
(symbols) and to the EPOS model~\cite{EPOS} predictions (dashed
lines). The values obtained from experimental data and from EPOS
predictions are both taken from
Ref.~\cite{NA61_pp}.}\label{inverse_slope}
\end{figure*}

The values of $T$ parameters fitted by the authors of
Ref.~\cite{NA61_pp} to EPOS simulated spectra seem to provide a fair
agreement with those extracted therein from experimental data, at
least for negative pions, positive pions starting from 40 GeV/$c$
beam momentum, and negative kaons. The situation is more complicated
for the values of $T$ extracted from fits to our UrQMD simulation.

For positive pions, the UrQMD-based inverse slope $T$ gives a
remarkably good agreement with that extracted from experimental
data, and describes well the characteristic decrease of the latter
with increasing pion rapidity. Being representative of the first
order characteristics of the {\em shape} of the $p_T$ distribution,
the inverse slope predicted by the UrQMD matches the energy
dependence seen in experimental data better than the
$d^2N/dydp_T(y,p_T)$ distribution itself as shown in
Figs~\ref{dNdptdy_pion_y=0.1} and ~\ref{dNdptdy_pion_y=1.5} (right),
and better than the $p_T$-integrated $dN/dy$ spectrum and mean
$\pi^+$ multiplicity discussed above.

The situation is less favorable for negative pions, although the
agreement between the values extracted from the experimental data
and the UrQMD is satisfactory at top SPS energy. With decreasing
beam momentum, the $\pi^-$ inverse slope at forward rapidity takes
off from that extracted from experimental data, reaching deviations
of about 20\% in the lower energy regime where also significant
discrepancies were visible for the mean $\pi^-$ multiplicity and the
$dN/dy(y\approx 0)$ yield as mentioned above. We note that the
agreement between data and model remains good for all collision
energies in the central region, near $y\approx 0$.

For positively charged kaons, a similarly good agreement is present
between NA61/SHINE data and UrQMD for the $T$ parameter extracted at
central rapidity for all energies, while a take-off of the UrQMD
prediction from experimental data at high rapidity is apparent at
all beam momenta apart from $p_{lab}=40$~GeV/$c$. This is somewhat
in contrast with the energy dependence of the  $d^2N/dydp_T(y,p_T)$
distribution, Figs~\ref{dNdptdy_Kaon_y=0.1}
and~\ref{dNdptdy_Kaon_y=1.5} (right), the $dN/dy$ spectrum, and the
mean $K^+$ multiplicity, where the discrepancies between data and
model gradually increased with decreasing energy.

Finally, for negatively charged kaons, the values of $T$ fitted to
the experimental data in Ref.~\cite{NA61_pp} seem to suffer from
fluctuations especially in the lower beam momentum regime. This
taken into account, we conclude that the overall description of the
rapidity dependence of the inverse slope as function of collision
energy is reproduced by the UrQMD reasonably well. This is
notwithstanding the discrepancies between experimental data and
model in terms of the energy dependence of the forward rapidity
$d^2N/dydp_T(y,p_T)$ distribution, Fig.~\ref{dNdptdy_Kaon_y=1.5}
(left), or the mean multiplicity of $K^-$ mesons as discussed above.

\section{Summary and conclusions}
\label{Summary} In the present paper we analyzed the new data on
proton, antiproton and meson production in $p+p$ collisions in the
laboratory momentum range of 20 to 158 GeV/$c$ recently published by
the NA61/SHINE Collaboration~\cite{NA61_pp}. These new experimental
results were compared to simulations performed using the recent
version 3.4 of the UrQMD transport model. Complementary to our
earlier analysis of mean multiplicities and $dN/dy$ distributions
published in Ref.~\cite{NA61_pp}, we focused on the transverse
momentum spectra of $\pi^{\pm}$, $K^{\pm}$, $p$ and $\bar p$
produced at central and forward rapidity, starting from beam energy
of 31 GeV. We also fitted these spectra with the same analytical
parametrization as applied to experimental data and compared the
respective inverse slopes as a function of meson rapidity and
collision energy.

Taken together the information on mean multiplicities, rapidity
distributions and $d^2N/dydp_T$ spectra, an overall discrepancy
between the UrQMD and the NA61/SHINE antiproton data is apparent.
The model overestimates antiproton production at all collision
energies where experimental data are available. For protons, a
discrepancy between the experimental data and the model can be seen
in the transport of baryon number down to low proton rapidity, in
particular at top SPS energy. Also the $p_T$ spectra of protons
simulated by the UrQMD cannot be reasonably fitted with
parametrization~(\ref{fit_function}), and differ from experimental
data especially at higher beam momenta.

For charged pion production in $p+p$ collisions, the fair
description provided by the UrQMD at higher SPS energies (80-158
GeV) considerably deteriorates with decreasing collision energy, to
a different extent for the three observables mentioned above. A
peculiar exception to this rule is the rapidity dependence of the
fitted $\pi^+$ inverse slope, which describes the data surprisingly
well down to the lowest considered beam momentum of 31~GeV/c.

For strange $K$ mesons a complicated pattern of discrepancies
between data and model is apparent, the most significant being the
fact that the UrQMD significantly underpredicts positive kaon
production in the lower SPS energy regime.

In view of the importance of the SPS (and RHIC beam energy scan)
energy regime which is claimed to host the onset of deconfinement
from hadronic matter to quark-gluon plasma in heavy ion
collisions~\cite{NA49_deconfinement_1,NA49_deconfinement_2,smes},
the significance of reference $p+p$ collisions cannot be stressed
enough. This is even more evident in view of the apparent
similarities in the energy dependence of kaon inverse slopes and
kaon-over-pion ratios in $p+p$ and heavy ion collisions, reported by
the NA61/SHINE Collaboration~\cite{Aduszkiewicz-qm2017}. In this
situation, the application of transport models is particularly
valuable as it gives a chance to follow in detail whether these
similarities have a decisive or only a casual importance for our
knowledge of conditions proper for the formation of deconfined
quark-gluon plasma matter.

In the above context, however, our analysis shows that the new
experimental data from NA61/SHINE constitute still a challenge for
specific transport models, at least as far as the present 3.4
version of the UrQMD code is concerned. As we mentioned in
Introduction the possible improvements to the UrQMD model were
considered for instance in Ref.~\cite{Uzh}. In particular, the
$\eta$-meson decays were implemented to the model as an additional
source of $\pi$-meson and $K$-meson production in inelastic $p+p$
interactions at SPS energies. However, it was shown by the author
that accounting of $\eta$-meson decays in the simulations leads to
the increase of the spectra of pions in central regions only by a
few percent and therefore it has almost a negligible effect on
$K$-meson production. This clearly is not sufficient for a good
description of the experimental data. On the other hand, it is
argued that decreasing of the cross sections of binary inelastic
reactions and accurate parameterizations of single diffraction cross
sections would allow the UrQMD model to describe meson production in
$p+p$ collisions (see Fig.~3 in Ref.~\cite{Uzh}). As stated in the
cited reference, in order to improve baryon production the inclusion
of the low mass diffraction dissociation to the UrQMD model could be
performed by increasing the cross section for the process
$NN\rightarrow\Delta N^*$, including the states $N^*$ with masses in
the range $m=1440,\ldots,2250$~MeV/$c^2$. To sum up, these and other
improvements could be tested in view of a possible better
description of the new available NA61/SHINE data on the energy
dependence of $p+p$ interactions. However, account taken that heavy
ion collisions have up to now the prime interest of the model, what
is to be envisaged is a complex tuning process where not only $p+p$,
but also $A+A$ collisions would come into the game. It is therefore
to be considered whether the new data on $Be+Be$ and $Ar+Sc$
collisions presently in progress from the NA61/SHINE Collaboration
(see preliminary data in~\cite{Aduszkiewicz-qm2017}) would not be
necessary to obtain a robust state-of-the-art description of
particle production at SPS energies.

%One can summarize that improvements discussed above can be
%implemented to the UrQMD model for better description of the
%experimental data on particle production in $p+p$ interactions at
%SPS energies, however, it is important to mention here that the
%effect of these changes on the description of the heavy-ion
%collisions, for what the UrQMD model is mostly dedicated, must be
%studied as well.

%A possible discrimination between the different model assumptions,
%as well as a detailed tuning of model parameters seems therefore
%highly indicated as a new step towards a better understanding of the
%strong interaction at high energies.

\section*{Acknowledgments}
The authors warmly thank Jan Steinheimer for his guidance with the
UrQMD code, Marek Gazdzicki for his inspiring remarks, and Seweryn
Kowalski as well as Szymon Pu\l{}awski for their help with the
experimental results from NA61/SHINE. This work was supported by the
National Science Centre, Poland under Grant No. 2014/14/E/ST2/00018.

\end{document}